\documentclass[lettersize,journal]{IEEEtran}
\usepackage{amsmath,amsfonts}
\usepackage{algorithmic}
\usepackage{algorithm}
\usepackage{array}
\usepackage[caption=false,font=normalsize,labelfont=sf,textfont=sf]{subfig}
\usepackage{textcomp}
\usepackage{stfloats}
\usepackage{url}
\usepackage{verbatim}
\usepackage{graphicx}
\usepackage{cite}
\usepackage[normalem]{ulem}
\usepackage{graphicx}
\usepackage{float}
\usepackage[utf8]{inputenc}
\usepackage{hyperref}
\usepackage{array}
\usepackage{multirow}
\usepackage[dvipsnames]{xcolor}
\hypersetup{colorlinks=true, linkcolor=blue, citecolor=blue, filecolor=magenta, urlcolor=cyan}

\begin{document}

\title{Routes for light management in monolithic perovskite/silicon tandem solar cells}

\author{Michel G. Rocha and Emiliano R. Martins%
\thanks{The authors are with the Department of Electrical and Computer Engineering, São Carlos School of Engineering, University of São Paulo, São Carlos, SP 13566-590, Brazil (e-mail: michelgr@usp.br; erm@usp.br).}
}

\maketitle

\markboth{IEEE Photonics Journal}{Routes for light management in monolithic perovskite/silicon tandem solar cells}

\begin{abstract}
Fully-textured perovskite/silicon tandem solar cells have emerged as promising candidates for next-generation photovoltaics. The optical functions of full texturing, however, are not yet fully understood. A key challenge is the requirement for perovskite layer texturing, which often leads to increased electrical losses. Here, we elucidate the distinct optical roles of front and rear textures in tandem configurations using optical simulations and use these insights to propose a new architecture that eliminates the need for perovskite surface texturing. We demonstrate that our proposed structure achieves optical results comparable to those of fully-textured devices, while its planar perovskite layer has the potential to reduce electrical losses. The high optical performance also results in higher efficiency if a texture-induced voltage loss as low as 50 mV is assumed, which is about six times lower than the loss of fully-textured devices, thus enabling higher efficiencies within a simplified design. Our results show that perovskite texturing is not essential for optimal light management, thus opening the way to combine efficient light management with high electrical performance.
\end{abstract}


\section{Introduction}

\IEEEPARstart{A}{s} the efficiency of single-junction silicon solar cells approaches its theoretical limit~\cite{GREEN} ($\sim$30\%~\cite{SHOCKLEY}), increasing attention has turned to tandem solar cells (TSCs), which combine materials with distinct band gaps within a single device. Certified power conversion efficiencies (PCEs) of up to 34.6\% have already been demonstrated~\cite{NREL}, and their theoretical limits can be as high as 45.2\%~\cite{VOS,RAJI}.

\par Tandems based on a perovskite top cell and a silicon bottom cell hold great promise for low-cost, high-efficiency devices. Perovskites exhibit tunable band gaps~\cite{NOH}, strong absorptance in the visible and near-UV spectrum~\cite{ALBRECHT}, and carrier diffusion lengths exceeding their absorption depth~\cite{LIANG,SAMUEL}. Perovskite materials are also inexpensive and composed of earth-abundant elements~\cite{CHUNG}. When combined with the low band gap of silicon, they enable broad-band light harvesting at low manufacturing costs~\cite{BAILIE}.

\par Achieving high PCEs in TSCs, however, requires effective light management to direct the appropriate portions of the solar spectrum to each sub-cell~\cite{WHITE}, enhancing absorptance~\cite{LAL,LI2} and minimizing reflection losses~\cite{YU,RAZA}. This can be accomplished through photonic structures placed at intermediate interfaces~\cite{LAL,QIAOJING,JACOBS,SANTBERGEN} or via optical impedance matching layers~\cite{MARTINS}. Various light-trapping and anti-reflection strategies have been proposed for perovskite/silicon tandem solar cells (PSTSCs), such as Mie-resonance-based spectral splitting~\cite{POLMAN}, nanosphere~\cite{EISENLOHR} and nanohole-array~\cite{KUANG} diffractive structures, planar black silicon textures~\cite{ARRUDA}, and light management foils~\cite{JOST}. Among these, pyramidal texturing has emerged as a particularly effective solution, providing efficient light trapping and reducing reflectance~\cite{SHI, SCHNEIDER, FOSTER, SAHLI, ALASHOURI, TOCKHORN}.


\par The impressive optical gains achieved through pyramidal texturing in single-junction silicon cells have inspired extensive research on textured PSTSCs~\cite{CHEN, AYDIN, HOU, AYDIN2, CHIN, KORE, MARIOTTI}. However, these textures introduce additional surface area and fabrication complexity, often resulting in elevated electrical losses that offset optical improvements~\cite{RAJI, CHEN, WANG, SHIREV}. Consequently, the experimental efficiencies remain significantly lower than the theoretical predictions~\cite{SHI, SCHNEIDER, FOSTER, VOS}.

\par Fully-textured PSTSCs — such as the structure illustrated in Figure~\ref{fig-tandem}(a) — have emerged as the most promising architecture for light management, with reported PCEs among the highest to date~\cite{LONGI,MARIOTTI,KORE,ALASHOURI}. The roles of anti-reflection and light trapping in such structures, however, are not yet fully understood. Moreover, there are significant challenges associated with full texturization, particularly regarding conformal deposition over steep textures and the resulting non-radiative recombination losses in the perovskite layer~\cite{FU,SHIREV}, which can exceed 300~mV~\cite{RAJI}. In particular, the benefit of texturing the perovskite interface remains controversial, as electrical losses may counteract the optical gains~\cite{WANG}. Furthermore, planar perovskite layers are increasingly regarded as a more reliable pathway toward scalable, high-quality, large-area cells~\cite{LUO,AYDIN3}.

\par Here, we elucidate the optical role of pyramidal texturing in PSTSCs and show that effective light management can be achieved without perovskite texturing. We then use these insights to propose a simplified architecture that achieves comparable optical performance without requiring perovskite texturing. We show that the proposed design can outperform the efficiency of fully-textured PSTSCs when texture-induced voltage losses as low as 50 mV are assumed, which is at least six times lower than losses reported in the literature ~\cite{RAJI}. Our results show that perovskite texturing is not essential for optimum light management, thus opening the way for tandem cells with efficient light management and high electrical performance.

\section{Structures, Parameters and Methods}\label{sec-methods}

\par
\begin{figure}[ht]
    \centering
    \includegraphics[scale=0.17]{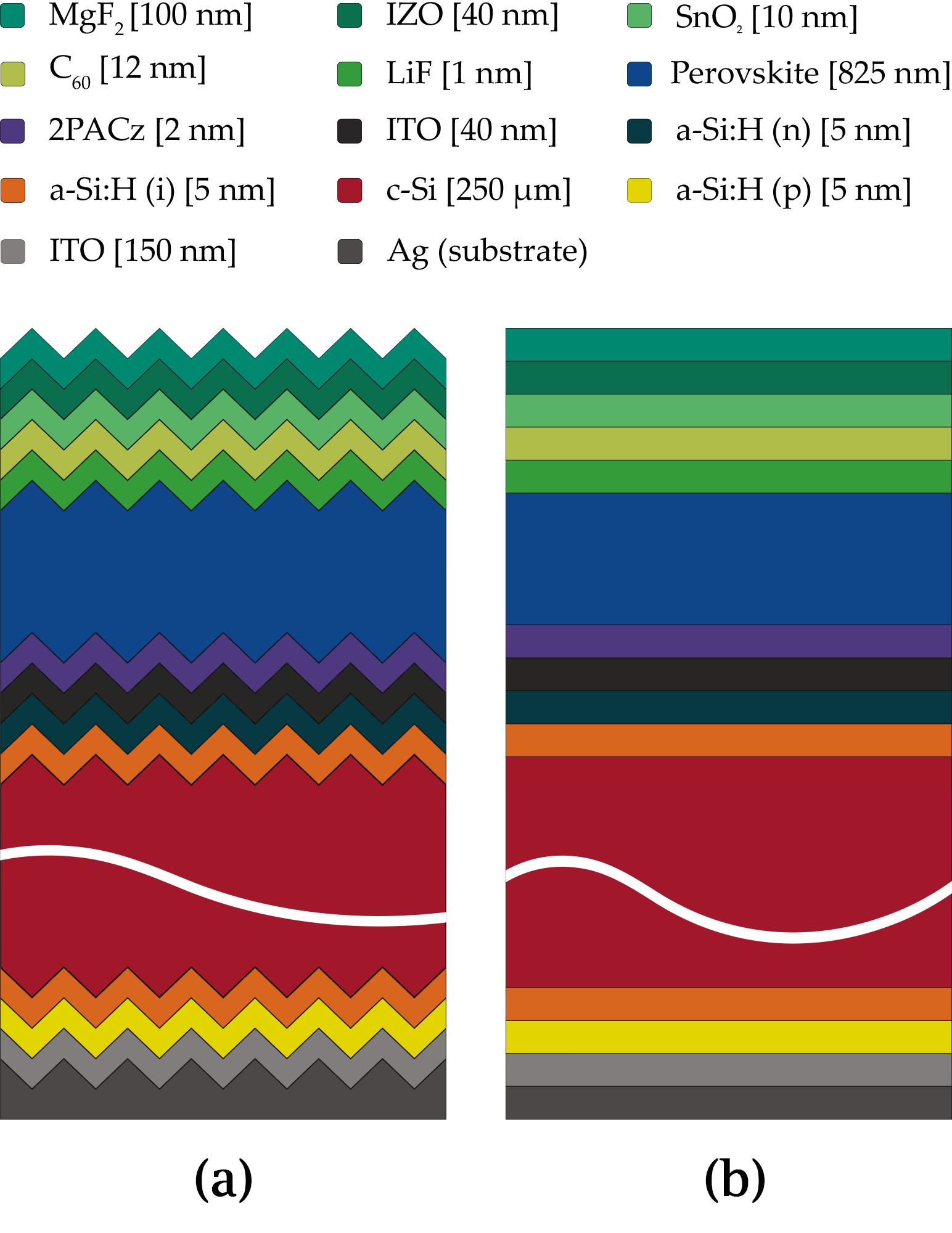}
    \caption{Tandem cell architectures used for optical benchmarking: (a) fully-textured, state-of-the-art configuration; (b) planar reference.}
    \label{fig-tandem}
\end{figure}

 To elucidate the roles played by full texturing, we consider the tandem solar cell of Figure \ref{fig-tandem}(a), which is representative of fully-textured cells \cite{RAJI,CHEN,CHIN,SAHLI}. A planar stack, shown in Figure \ref{fig-tandem}(b), is also considered as a reference. The optical simulations are based on Rayflare \cite{RAYFLARE}, which employs a multimodal approach to model optical structures combining geometric and wave optics. The pyramidal textures in Figure \ref{fig-tandem} are modeled using a Phong scattering profile, which accounts for surface roughness and diffuse reflection~\cite{FUNG}. The electrical properties of the cells are modeled by a two-diode equivalent circuit through Solcore \cite{SOLCORE}. We assume typical values for saturation currents $J_{0_1}$ and $J_{0_2}$, ideality factors $n_1$ and $n_2$, series resistance $R_\mathrm{s}$ and shunt resistance $R_\mathrm{sh}$~\cite{HAODIODE,ZENG,KIERMASCH}, as summarized in Table \ref{tab-twodiode}. 

\par The short-circuit current density ($J_{\mathrm{sc}}$) of each subcell is calculated according to Eq.~\ref{eq-jsc}, where $\phi_{\mathrm{AM1.5G}}$ is the AM1.5G solar spectrum, $f_\mathrm{c}$ is the collection efficiency of photogenerated carriers and $A(\lambda)$ is the spectral absorptance. The collection efficiency for perovskite is determined using Eqs.~\ref{eq-fc} and~\ref{eq-lambda}~\cite{WHITE}, where $k$ is the Boltzmann constant, $T$ is the device operating temperature (298~K), $W$ is the layer thickness, $L_\mathrm{d}$ is the minority carrier diffusion length (assumed to be 1~$\mu$m~\cite{LIANG}), and V$_{\mathrm{bi}}$ is the built-in voltage, assumed equal to the perovskite band gap. The refractive indices of all materials are extracted from Refs.~\cite{PVLIGHTHOUSE,RIINFO,FUJIWARA2} (see Supplementary Information I, Figure S1). The optical data for the perovskite layer assumes a Cs$_{0.17}$FA$_{0.83}$Pb(I$_{0.83}$Br$_{0.17}$)$_3$ composition~\cite{FUJIWARA2}, with a band gap of 1.63~eV.

\begin{equation}\label{eq-jsc}
    J_{\mathrm{sc}} = f_c q \int_{320\,\mathrm{nm}}^{1200\,\mathrm{nm}} \phi_{\mathrm{AM1.5G}}(\lambda) A(\lambda) \, d\lambda
\end{equation}

\begin{equation}\label{eq-fc}
    f_c = \frac{e^{\lambda_2^0/2}-1}{\lambda_2^0/2}
\end{equation}

\begin{equation}\label{eq-lambda}
    \lambda_2^0 = \frac{q V_{\mathrm{bi}}}{2kT} - \sqrt{\left( \frac{W}{L_d} \right)^2 + \left( \frac{q V_{\mathrm{bi}}}{2kT} \right)^2}
\end{equation}

\par These parameters are then used to obtain the $J$–$V$ characteristics of each subcell based on a two-diode model.

\begin{table}[!t]
\centering
\caption{Two-diode model parameters for the silicon and perovskite solar subcells}
\label{tab-twodiode}
\renewcommand{\arraystretch}{1.5}
\setlength{\tabcolsep}{4pt}  
\footnotesize  
\begin{tabular}{lcccccc}
\hline
\textbf{Solar} & $J_{0_1}$ & $J_{0_2}$ & $n_1$ & $n_2$ & $R_\mathrm{s}$ & $R_\mathrm{sh}$ \\
\textbf{Subcell} & (mA/cm$^2$) & (mA/cm$^2$) & - & - & ($\Omega\cdot$cm$^2$) & ($\Omega\cdot$cm$^2$) \\
\hline
Perovskite & $3 \times 10^{-18}$ & $9 \times 10^{-10}$ & 1.3 & 2.0 & 0.4 & $3 \times 10^3$ \\
Silicon    & $6 \times 10^{-11}$ & $5 \times 10^{-6}$  & 1.2  & 2.0 & 0.2 & $1 \times 10^5$ \\
\hline
\end{tabular}
\end{table}

\section{Optical Role of Full Texturing}\label{sec-results}

\par
\begin{figure}[ht]
    \centering
    \includegraphics[scale=0.3]{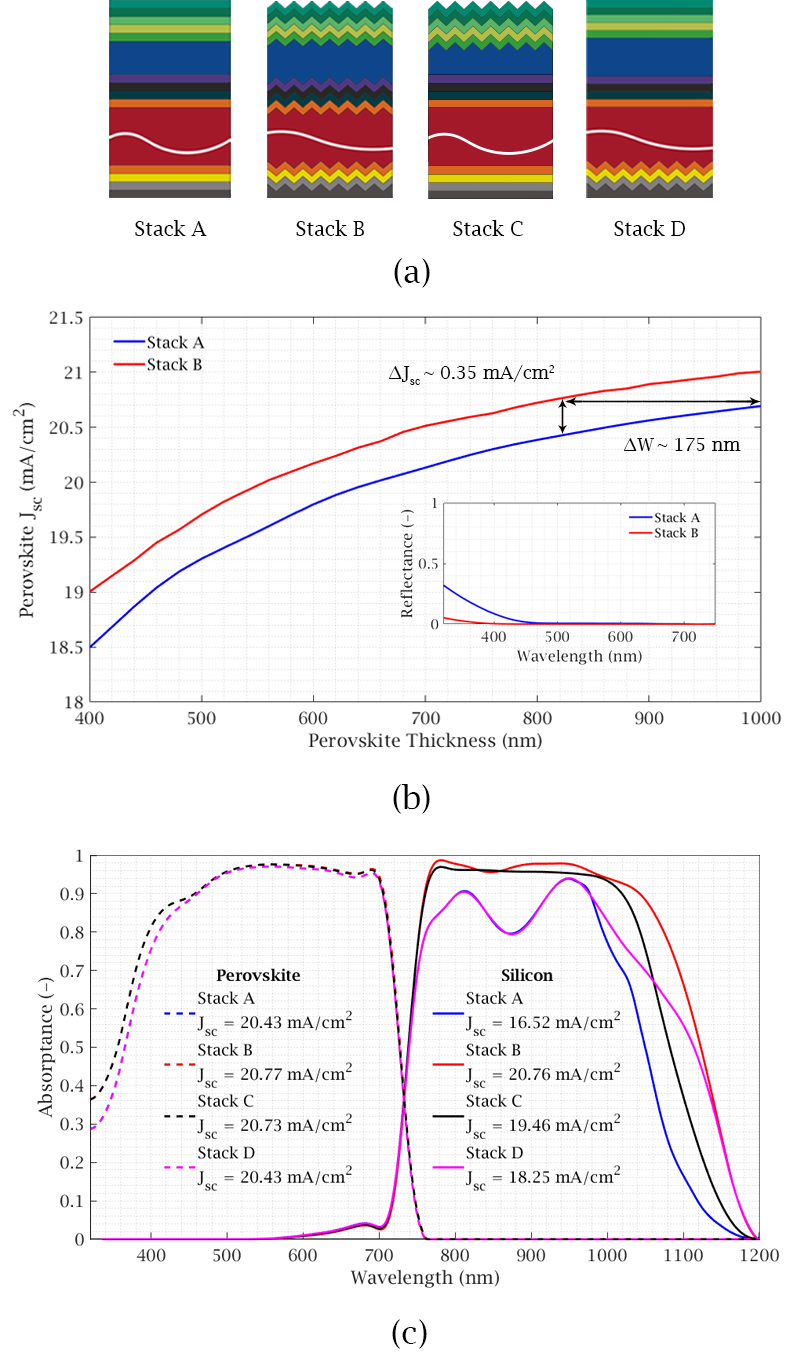}
    \caption{Light management effects in a fully-textured PSTSC architecture: (a) device stacks considered for the optical characterizations; (b) short-circuit current density ($J_\mathrm{sc}$)  of the perovskite subcell for planar (stack A) and fully-textured (stack B) structures; inset: front-surface reflectance from stacks A and B; (c) spectral absorptance in the perovskite and silicon layers for stacks A--D.}
    \label{fig-arlt}
\end{figure}

 Figure \ref{fig-arlt} shows the reference stack A, the fully-textured stack B, and two additional stacks (C and D) used to help gain insight into the optical roles of front and rear texturing. All stacks have identical active layer thicknesses and optical properties, as outlined in Figure~\ref{fig-tandem}.

\par Figure~\ref{fig-arlt}(b) shows the short-circuit current density ($J_{\mathrm{sc}}$) of the perovskite subcell as a function of its thickness for a planar (stack A — blue line) and fully-textured configuration (stack B — red line). The vertical arrow highlights the short-circuit current density enhancement in a typical tandem device with a 250~$\mu$m-thick silicon subcell \cite{AYDIN}, for which current matching is achieved with a perovskite thickness of 825~nm. Notice that the same current enhancement can be achieved simply by increasing the planar perovskite thickness to $\sim$1000~nm, which is within practical fabrication limits and comparable to carrier diffusion lengths. As the necessity of current matching in tandem cells fixes an upper bound for the useful perovskite current (here the upper bound is $\sim$20.8 mA/cm$^2$, which is close to the best state-of-the-art values obtained~\cite{GREEN,AYDIN}), it follows that the perovskite matching current can be achieved with thicker planar perovskite layers, with no need for light trapping, unless the silicon layer thickness is significantly increased beyond its typical values. This feature can be attributed to the combination of high absorption of perovskites and their long carrier diffusion lengths, which exceed 1~$\mu$m~\cite{LIANG,SAMUEL} -- check Supplementary Information II, Figures S2 and S3. 

\par If light trapping does not play an essential role in perovskite absorption, then the question of the role of the front surface texturing arises. One possibility is that front texturing acts as an anti-reflection layer; however, interestingly, although pyramids have excellent anti-reflection properties, their performance in PSTSCs is not significantly superior to a planar AR coating (stack A), as shown in the inset of Figure~\ref{fig-arlt}(c), which compares the reflectance from stacks A and B.

\par The role of front texturing emerges when we compare the silicon absorption spectra of stacks A, B, C and D (Figure~\ref{fig-arlt}(c)). Note that the absorption spectra of stack A (blue line) and D (magenta line), which do not employ front texturing, are similar in the wavelength range between $\sim$700 nm and $\sim$980~nm. The superior absorption of stacks B (red line) and C (black line) in this spectral region indicates the silicon light-trapping role of the front texture. In particular, note the two peaks at $\sim$800~nm and $\sim$950~nm for stacks A and D, which disappear for stacks B and C. These peaks arise from thin-film interference in the perovskite layer, and their vanishing in stacks B and C is consistent with the stronger scattering imposed by front texturing in the wavelength range between $\sim$700 and $\sim$980~nm. The role of rear texturing, on the other hand, is to trap light in the longer wavelength region (above 1000 nm), as seen by comparing the absorption of stacks C and D.

\section{Proposed Structure}\label{sec-ps}

\par
\begin{figure}[ht]
    \centering
    \includegraphics[scale=0.23]{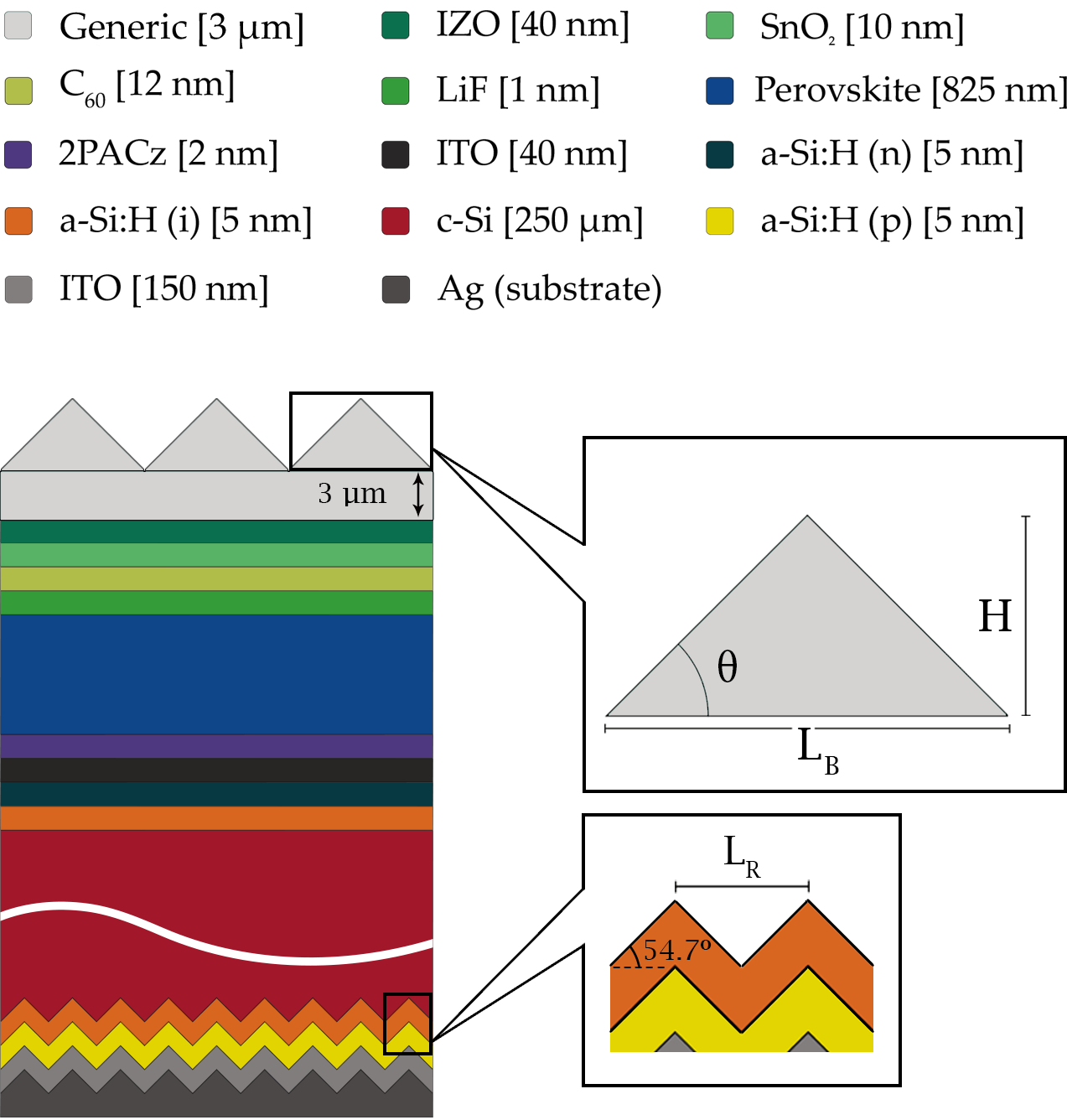}
    \caption{Proposed structure. The top inset highlights the geometry of the anti-reflective front-side texture optimized here. The bottom inset shows the state-of-the-art pyramidal texturing used at the rear silicon interface, which remains unaltered in comparison to all previous textured stacks.}
    \label{fig-ps}
\end{figure}

 Based on these insights, we propose a new light management scheme in which the perovskite layer remains planar, as illustrated in Figure~\ref{fig-ps}. This scheme aims to optimize silicon absorption by combining front and rear texturing -- the former targeting the $\sim$700 nm to $\sim$980~nm region and the latter targeting the region above 1000 nm. The silicon texturing follows conventional state-of-the-art parameters to ensure alignment with existing manufacturing techniques \cite{TIAN}.

\par The anti-reflective parameters described in Figure~\ref{fig-ps} were obtained by optimizing the inclination angle $\theta$ of the front-side pyramids. Our simulations show that the scattering from the ARC pyramids is not fully diffuse. The height $H$ of these pyramids was set to 3~$\mu$m to ensure compatibility with standard fabrication methods such as photolithography or laser interference lithography, which favor scalability~\cite{TRAUB}. A planar slab of the same material and thickness was placed under the texture as a buffer protecting the transport layers. We also assume that the front-side texture has a generic material with refractive index of $n_r = 1.4$, which is representative of the refractive indexes of materials commonly used in lithography, such as polydimethylsiloxane (PDMS)~\cite{TRAUB,ZYING}. The remaining layers follow the same architecture as Stack D in Figure \ref{fig-arlt} (a), consisting of a 40 nm indium zinc oxide (IZO) followed by 10 nm of tin oxide (SnO$_2$), a 12 nm of buckminsterfullerene (C$_{60}$) and a 1 nm lithium fluoride (LiF) interlayer. The 825 nm perovskite absorber is followed by a 2 nm layer of 2-(9H-carbazol-9-yl)ethylphosphonic acid (2PACz) and a 40 nm indium tin oxide (ITO) electrode, featuring the p-i-n architecture of the top perovskite subcell. As for the silicon subcell placed immediately below, it consists of a 5 nm n-doped hydrogenated amorphous silicon (a-Si:H (n$^{+}$)) layer followed by a 5 nm intrinsic a-Si:H (i) layer. The 250 µm crystalline silicon absorber is then followed by a 5 nm rear intrinsic a-Si:H (i) layer and a 5 nm p-doped a-Si:H (p$^{+}$) layer, forming a silicon heterojunction (SHJ). A 150 nm ITO electrode is deposited underneath followed by a silver (Ag) metal contact. The complete stack is depicted in Figure \ref{fig-ps}.
\par
\begin{figure}[ht]
    \centering
    \includegraphics[scale=0.31]{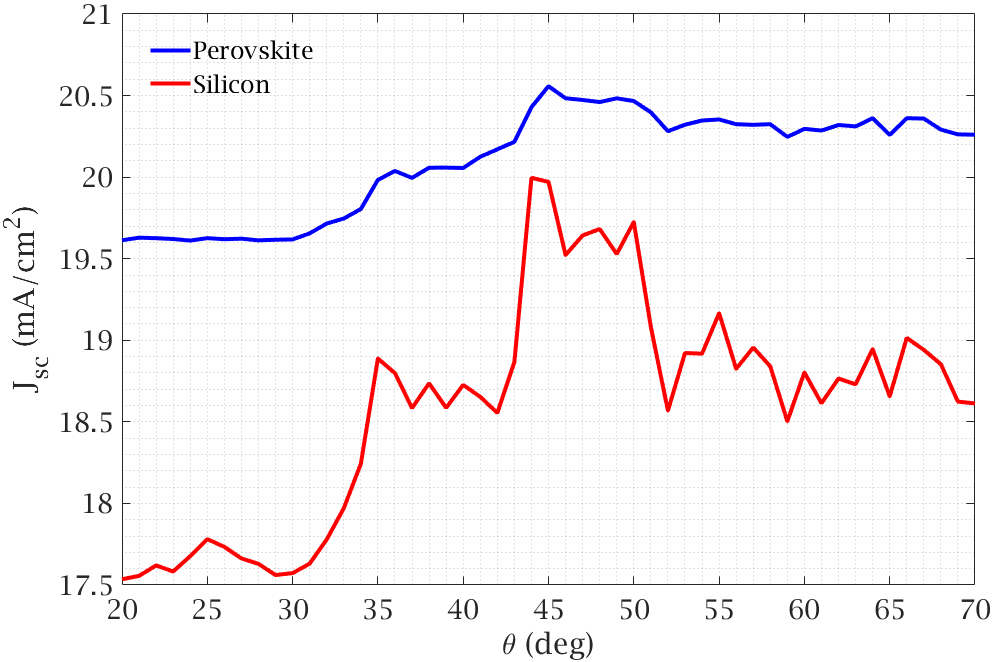}
    \caption{Short-circuit current density ($J_{\mathrm{sc}}$) as a function of the ARC pyramid inclination angle. The optimal angle of 44$^\circ$ maximizes both the silicon and perovskite currents.}
    \label{fig-angular}
\end{figure}

 The inclination angle of the front-side pyramids, $\theta$, was varied in the 20--70$^\circ$ range to maximize the short-circuit current density of silicon. As shown in Figure~\ref{fig-angular}, the silicon subcell reaches a maximum of $J_{\mathrm{sc}}$ at the optimum angle of $\theta = 44^\circ$. This angle entails a pyramid base length of $L_\mathrm{B} \simeq$ 6.2~$\mu$m. Interestingly, the silicon and perovskite currents are optimized by similar angles. 

\par
\begin{figure}[ht]
    \centering
    \includegraphics[scale=0.32]{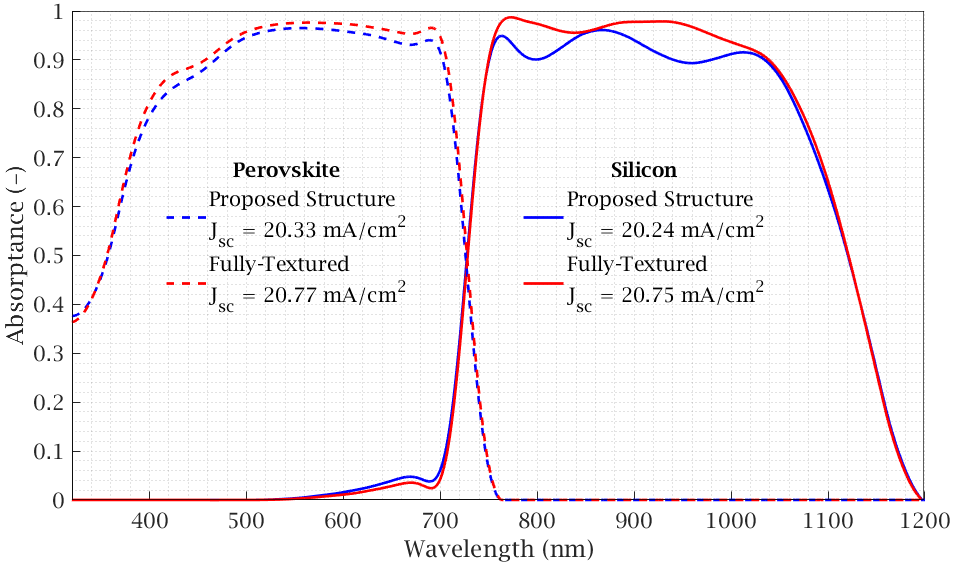}
    \caption{Spectral absorptance in the perovskite and silicon subcells for the proposed and fully-textured architectures. The proposed structure maintains comparable optical performance while simplifying the top interface.}
    \label{fig-psvsfull}
\end{figure}

\begin{figure}[ht]
    \centering
    \includegraphics[scale=0.29]{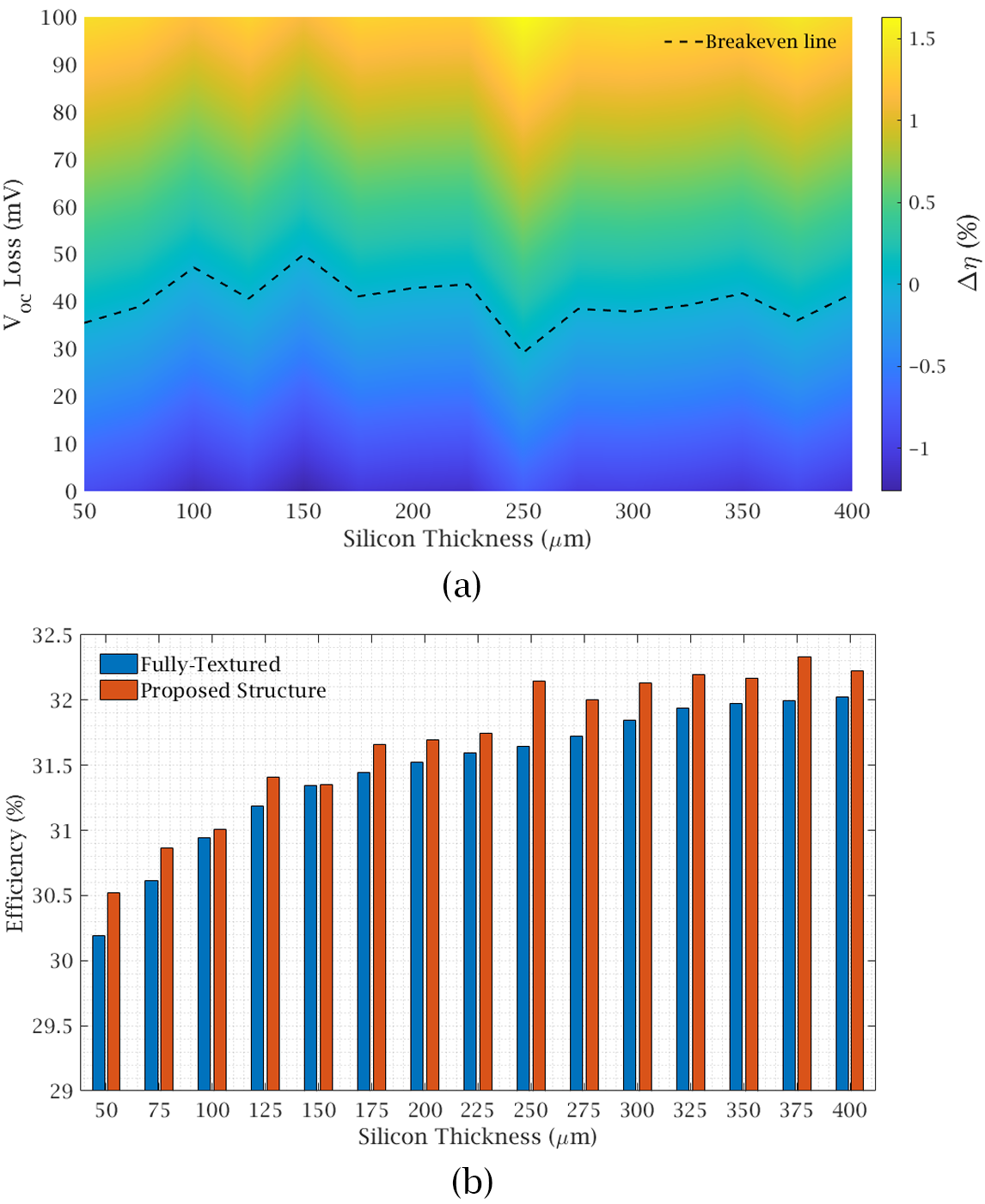}
    \caption{Performance comparison between the proposed and fully-textured architectures; (a) difference between the efficiencies of the proposed and fully-textured architectures; the black dashed line marks the boundary where the efficiencies are equal; (b) absolute efficiency comparison between both architectures when a 50~mV V$_\mathrm{oc}$ drop is considered for the fully-textured cell.}
    \label{fig-effs}
\end{figure}

 The optical performance of the proposed structure is shown in Figure~\ref{fig-psvsfull} alongside the fully-textured structure for reference. Both spectra correspond to current-matched cells, with optimized efficiencies of 32.15\% and 32.94\% for our structure and the fully-textured reference, respectively. These efficiencies, however, do not take into account parasitic losses, which are higher in the fully-textured solar cell~\cite{SHIREV,RAJI,WANG}. An estimate of the impact of losses arising from perovskite texturing is shown in Figure~\ref{fig-effs}(a), where $\Delta\eta$ is the absolute PCE difference between the proposed and fully-textured architectures, respectively. The dashed line, which peaks at around $\sim$50~mV, marks the boundary where our structure outperforms the fully-textured solar cell. Conformal perovskite deposition on textured substrates has been shown to lead to severe interfacial recombination both at the electron transport layer (e.g., C$_{60}$)~\cite{RAJI} and the hole transport layer (e.g. 2PACz)~\cite{WANG} interfaces, with experimental V$_\mathrm{oc}$ losses between 100~mV~\cite{SHIREV,FU} and 300~mV~\cite{RAJI}. Thus, these results point to the benefit of light trapping schemes in which the perovskite layer is kept planar, such as the scheme proposed here. Indeed, taking these losses into account\footnote{To account for the drop in V$_\mathrm{oc}$ in the fully-textured cell, the ideality factor $n_2$ of perovskite was varied in the range of 2--2.3, with the corresponding recombination current $J_{0_2}$ varying in the order of $10^{-10}$ to $10^{-8}$ mA/cm$^2$~\cite{KIERMASCH,ZENG}.}, our proposed structure has the potential to outperform current fully-textured solar cells across a wide range of silicon thicknesses, as shown in Figure~\ref{fig-effs}(b), while also requiring thinner perovskite layers — see Supplementary Information III, Figure~S4.   

\par Finally it is worth noting that, while the precise size of the pyramidal textures is not critical~\cite{HOHN}, their inclination must be preserved, particularly for the ARC texture, as corroborated by Figure~\ref{fig-angular}, which shows that the inclination angle itself plays a crucial role in effective light redirection. 

\section{Conclusions}\label{sec-conclusions}

\par In conclusion, we demonstrated that both front and rear texturing are essential to increase absorption in the silicon layer of perovskite/silicon tandem solar cells, but are not essential for perovskite light trapping. This insight led to the proposal of a light management scheme using a planar perovskite layer, thus promoting better electrical properties. The optical performance of this new architecture was shown to be comparable to fully-textured solar cells, yielding comparable efficiencies when a V$_\mathrm{oc}$ penalty as low as 50~mV is assumed in the latter. Since these penalties can exceed 300~mV in fully-textured cells~\cite{RAJI}, we believe that our strategy can surpass the efficiency of current fully-textured designs and point to new directions for light management in tandem solar cells.

\section*{Acknowledgements}

The authors acknowledge financial support from  São Paulo Research Foundation – FAPESP (Grants.  2020/00619-4, 2021/06121-0) and the Brazilian National Council for Scientific and Technological Development – CNPq (307602/2021-4, 303820/2024-1). The Article Processing Charge (APC) for the publication of this research was funded by the Coordination for the Improvement of Higher Education Personnel (CAPES) – ROR identifier: 00x0ma614. For the purpose of open access, the authors have applied the Creative Commons CC BY license to any accepted version of the manuscript.

\bibliographystyle{IEEEtran}
\bibliography{refs}

\end{document}